\documentclass[apjl]{emulateapj}

\usepackage{apjfonts}
\usepackage{graphicx}
\usepackage{natbib}

\newcommand{\lsi}{\objectname[LSI +61 303]{LS~I~+61\degr303}}
\newcommand{\lsicap}{LS~I~+61\degr303}

\shorttitle{Fermi Observations of LS~I~+61\degr303}
\shortauthors{Abdo et al.}

\begin{document}

\title{Fermi LAT Observations of LS~I~+61\degr303: \\ First detection of an orbital modulation in GeV Gamma Rays}



\author{
A.~A.~Abdo\altaffilmark{2,3}, 
M.~Ackermann\altaffilmark{4}, 
M.~Ajello\altaffilmark{4}, 
W.~B.~Atwood\altaffilmark{5}, 
M.~Axelsson\altaffilmark{6,7}, 
L.~Baldini\altaffilmark{8}, 
J.~Ballet\altaffilmark{9}, 
G.~Barbiellini\altaffilmark{10,11}, 
D.~Bastieri\altaffilmark{12,13}, 
B.~M.~Baughman\altaffilmark{14}
% ...
}
% ...
\altaffiltext{2}{National Research Council Research Associate, National Academy of Sciences, Washington, DC 20001}
\altaffiltext{3}{Space Science Division, Naval Research Laboratory, Washington, DC 20375}
\altaffiltext{4}{W. W. Hansen Experimental Physics Laboratory, Kavli Institute for Particle Astrophysics and Cosmology, Department of Physics and SLAC National Accelerator Laboratory, Stanford University, Stanford, CA 94305}
\altaffiltext{5}{Santa Cruz Institute for Particle Physics, Department of Physics and Department of Astronomy and Astrophysics, University of California at Santa Cruz, Santa Cruz, CA 95064}
\altaffiltext{6}{The Oskar Klein Centre for Cosmo Particle Physics, AlbaNova, SE-106 91 Stockholm, Sweden}
\altaffiltext{7}{Department of Astronomy, Stockholm University, SE-106 91 Stockholm, Sweden}
\altaffiltext{8}{Istituto Nazionale di Fisica Nucleare, Sezione di Pisa, I-56127 Pisa, Italy}
\altaffiltext{9}{Laboratoire AIM, CEA-IRFU/CNRS/Universit\'e Paris Diderot, Service d'Astrophysique, CEA Saclay, 91191 Gif sur Yvette, France}
\altaffiltext{10}{Istituto Nazionale di Fisica Nucleare, Sezione di Trieste, I-34127 Trieste, Italy}
\altaffiltext{11}{Dipartimento di Fisica, Universit\`a di Trieste, I-34127 Trieste, Italy}
\altaffiltext{12}{Istituto Nazionale di Fisica Nucleare, Sezione di Padova, I-35131 Padova, Italy}
\altaffiltext{13}{Dipartimento di Fisica ``G. Galilei", Universit\`a di Padova, I-35131 Padova, Italy}
\altaffiltext{14}{Department of Physics, Center for Cosmology and Astro-Particle Physics, The Ohio State University, Columbus, OH 43210}
% ...

\begin{abstract}
This {\it Letter} presents the first results from the observations of
\lsi\ using Large Area Telescope data from the {\em Fermi Gamma-Ray
  Space Telescope} between 2008 August and 2009 March. Our results
indicate variability that is consistent with the binary period, with
the emission being modulated at $26.6 \pm 0.5$ days. This constitutes
the first detection of orbital periodicity in high-energy gamma rays
(20 MeV--100 GeV, HE). The light curve is characterized by a
broad peak after periastron, as well as a smaller peak just before
apastron.  The spectrum is best represented by a power law with an
exponential cutoff, yielding an overall flux above 100 MeV of 0.82
$\pm$ 0.03(stat) $\pm$ 0.07(syst) 10$^{-6}$ ph cm$^{-2}$ s$^{-1}$,
with a cutoff at 6.3 $\pm$ 1.1(stat) $\pm$ 0.4(syst) GeV and photon
index $\Gamma$ = 2.21 $\pm$ 0.04(stat) $\pm$ 0.06(syst).  There is no
significant spectral change with orbital phase. The phase of maximum
emission, close to periastron, hints at inverse Compton scattering as
the main radiation mechanism. However, previous very high-energy gamma
ray ($>$100 GeV, VHE) observations by MAGIC and VERITAS show peak
emission close to apastron. This and the energy cutoff seen with {\em
  Fermi} suggest the link between HE and VHE gamma rays is nontrivial.
\end{abstract}


\section{Introduction}

The high-mass X-ray binary \lsi\ (=\object{V615 Cas}) has long been
plausibly associated with a high-energy (HE, 20 MeV--100 GeV) gamma-ray
source, although never before confirmed. The discovery of the {\em COS
  B} source 2CG 135+01 \citep{1977Natur.269..494H} quickly brought
attention to this binary system's Be star localized within its error
box, because of its unusual periodic radio emission
\citep{1979AJ.....84.1030G} and its X-ray emission
\citep{1981ApJ...247L..85B}. 2CG 135+01 was to remain one of the
brightest sources known in the HE gamma-ray sky, with a flux of
$\sim 10^{-6}$~ph~s$^{-1}$~cm$^{-2}$ above 100 MeV
\citep{1981ApJ...243L..69S}. In the 1990s, EGRET detected the source
with high confidence at the same average flux level and derived a
power-law photon index of $\Gamma=2.05\pm0.06$
\citep{1997ApJ...486..126K}. Although there are no other objects of
note (radio-loud AGN, or pulsars) coinciding with the 3EG source
\citep{1999ApJS..123...79H}, its positional uncertainty was not small
enough to firmly associate the gamma-ray source with the binary.
Variability in the EGRET light curve could be neither firmly
established nor related to variability at other wavelengths
\citep{1998ApJ...497L..89T, 2003ApJ...597..615N}. Recently, {\em
  AGILE} has reported detecting the source at the same flux level
\citep{2009arXiv0902.2959P}.

\lsi\ is an unusual binary system exhibiting strong variable emission
from the radio to X-ray and TeV energies.  At radio wavelengths the
source has been shown to exhibit radio outbursts that are modulated on
an orbital period of 26.4960 $\pm$ 0.0028 days
\citep{1982ApJ...255..210T, 2002ApJ...575..427G}.  The phase of radio
maximum has also been shown by \citet{2002ApJ...575..427G} to vary
with a super-orbital period of 1667 $\pm$ 8 days. Observations of
orbital modulation in the optical place constraints on the
binary system parameters. The binary has an eccentric orbit
($e$=0.55-0.72) and the Be star radial velocity is consistent with a
neutron star companion or, if the orbital inclination is $\leq$
25\degr, with a $\geq 3 M_\odot$ black hole
\citep{1981PASP...93..486H,2005MNRAS.360.1105C}. Significant
uncertainty still exists in key parameters of the orbital solution of
the system \citep{2007ApJ...656..437G, 2009arXiv0902.4015A}.

Behavior in the X-ray band is much more complicated.  Orbital
modulation has been reported with the peak of emission appearing at
phases 0.6--0.7 \citep{1997A&A...320L..25P, 2007A&A...474..575E}.
However the modulation is not smooth, with short timescale flares and
very strong orbit-to-orbit variability \citep{2009ApJ...693.1621S}.
Broad-band spectral analysis of {\em XMM-Newton} and {\em INTEGRAL}
data by \citet{2006MNRAS.372.1585C} reveal \lsi\ to be well fitted by
a simple absorbed power-law with a hard photon index, $\Gamma
\simeq$1.5, in the 0.5--100 keV band.

The MAGIC telescope detected a variable very high-energy (VHE $>$100
GeV) gamma-ray source coincident with LS~I~+61\degr303
\citep{2006Sci...312.1771A}; a result that has been independently
confirmed by the VERITAS collaboration
\citep{2008ApJ...679.1427A}. More recently, the MAGIC collaboration
has further reported that the VHE emission is periodic at the 26.5 day
orbital period of the system \citep{2009ApJ...693..303A}. The VHE
emission is consistently highest close to apastron, when the compact
object is farthest from the Be star, and remains undetected at
periastron.  Like LS~5039 and PSR~B1259$-$63
\citep{2005Sci...309..746A,2005A&A...442....1A}, and contrary to Cyg
X$-$1 \citep{2007ApJ...665L..51A}, \lsi\ is a gamma-ray binary with
its spectral energy distribution peaking in HE gamma rays \citep[for a
  full SED see][]{2006A&A...456..801D, 2006MNRAS.372.1585C}.

Calculations of the theoretical expectations of the gamma ray emission
from \lsi\ go back almost three decades \citep{1981MNRAS.194P...1M},
and there has been a recent burst of activity following the MAGIC
detection.  Two scenarios have been put forward involving either the
relativistic wind of a young, rotation-powered pulsar
\citep{2006A&A...456..801D,2008APh....30..239S,2009ApJ...693.1462S},
or the relativistic jet of an accreting black hole or neutron star
\citep{2005ApJ...632.1093R,2006MNRAS.368..579B,2006ApJ...650L.123G,2006A&A...459L..25B}.
In light of the orbital modulations seen in radio, X-ray, and VHE
gamma rays, a detailed light curve in the HE gamma ray domain (where
most of the energy is output) is an essential piece to identify the
main radiative process at work and model the source.

\section{Data Reduction and results}

The \facility{{\em Fermi Gamma-ray Space Telescope}} was launched on
2008 June 11, from Cape Canaveral, Florida. The Large Area Telescope
(LAT) is an electron-positron pair production telescope, featuring
solid state silicon trackers and cesium iodide calorimeters, sensitive
to photons from $\sim 20$ MeV to $>300$ GeV
\citep{2009arXiv0902.1089F}. Relative to earlier gamma-ray missions
the LAT has a large $\sim 2.4$ sr field of view, a large effective
area ($\sim 8000$ cm$^2$ for $>$1 GeV on axis) and improved angular
resolution or point spread function (PSF, better than 1\degr\ for 68\%
containment at 1 GeV).  The {\em Fermi} survey mode operations began
on 2008 August 4, after the conclusion of a flawless commissioning
period.  In this mode, the observatory is rocked north and south on
alternate orbits to provide more uniform coverage so that every part
of the sky is observed for $\sim$30 minutes every 3 hours. Thus {\em
  Fermi} is ideally suited for long term all-sky observations. The
dataset for this analysis spanned 2008 Aug 4, through 2009 Mar
24. Thus \lsi\ was observed for approximately 9 orbital periods.

\begin{figure}
   \includegraphics[width=0.95\linewidth,angle=0,clip]{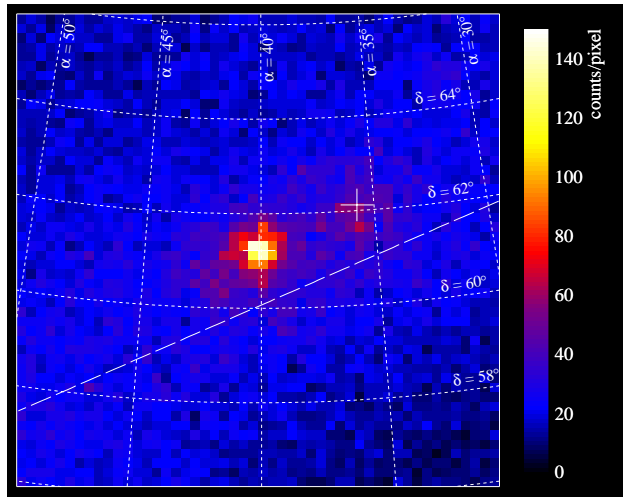}
\caption{The counts map for 100~MeV--300~GeV in (RA,Dec) of a
  10\degr\ region around the \lsicap\ location. The exposure varies by
  less than 2.5\% across the field at a representative energy of 10
  GeV.  The source is bright and fairly isolated, sitting on a
  background of Galactic and extragalactic diffuse emission. A fit to
  the source yields a significance of more than 70 $\sigma$. The
  dashed line indicates the Galactic equator ($b=$0); the crosses
  indicate the location of \lsicap\ (the brighter source) and a faint
  nearby point source.}
\label{CountsMap}
\end{figure}

The data were reduced and analysed using the {\em Fermi} Science Tools
v9r8 package\footnote{See the FSSC website for details of the Science
  Tools: http://fermi.gsfc.nasa.gov/ssc/data/analysis/}. The standard
onboard filtering, event reconstruction, and classification were
applied to the data \citep{2009arXiv0902.1089F}, and for this analysis
the high-quality ("diffuse") event class is used.Time periods when the
region around \lsi\ was observed at a zenith angle greater than
105\degr\ were also excluded to avoid contamination from Earth albedo
photons. With these cuts, a photon count map of a 10$^\circ$ region
around the binary is shown in Fig.~\ref{CountsMap}.  The alignment of
the LAT pointing direction with the celestial frame was calibrated
using a large set of high latitude gamma-ray sources to better than
10$\arcsec$ \citep{2009b.Adbo}.  The position of \lsi\ was found to be
R.A.~=~02$^{\rm h}$40$^{\rm m}$22\fs3,
Dec.~=~61\degr13\arcmin30\arcsec (J2000) with a 95\% error of
0.069\degr ; in agreement with the accepted position
\citep{2006smqw.confE..52D}.
 
 \subsection{Spectral Analysis}

\begin{figure}[t]
   \includegraphics[width=0.95\linewidth,angle=0,clip]{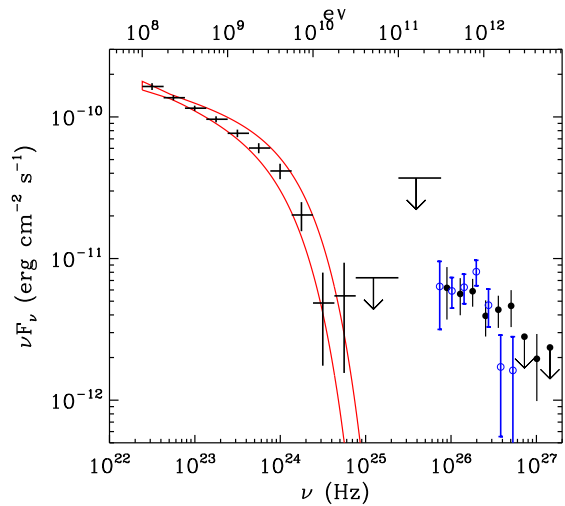}
\caption{Fitted spectrum of \lsicap\ to the phase-averaged {\em Fermi}
  data. The solid red lines are the $\pm$1$\sigma$ limits of the {\em
    Fermi} cutoff power law; blue (open circle) data points from MAGIC
  (high state phases 0.5--0.7); black (filled circle) data points from
  VERITAS (high state phases 0.5--0.8). Data points in the {\em Fermi}
  range are likelihood fits to photons in those energy bins. Note that
  the data from the different telescopes are not contemporaneous,
  though they do cover multiple orbital periods.}
\label{PhaseAveFit}
\end{figure}

\begin{figure*}[t]
\centering
\includegraphics[width=0.9\linewidth,clip]{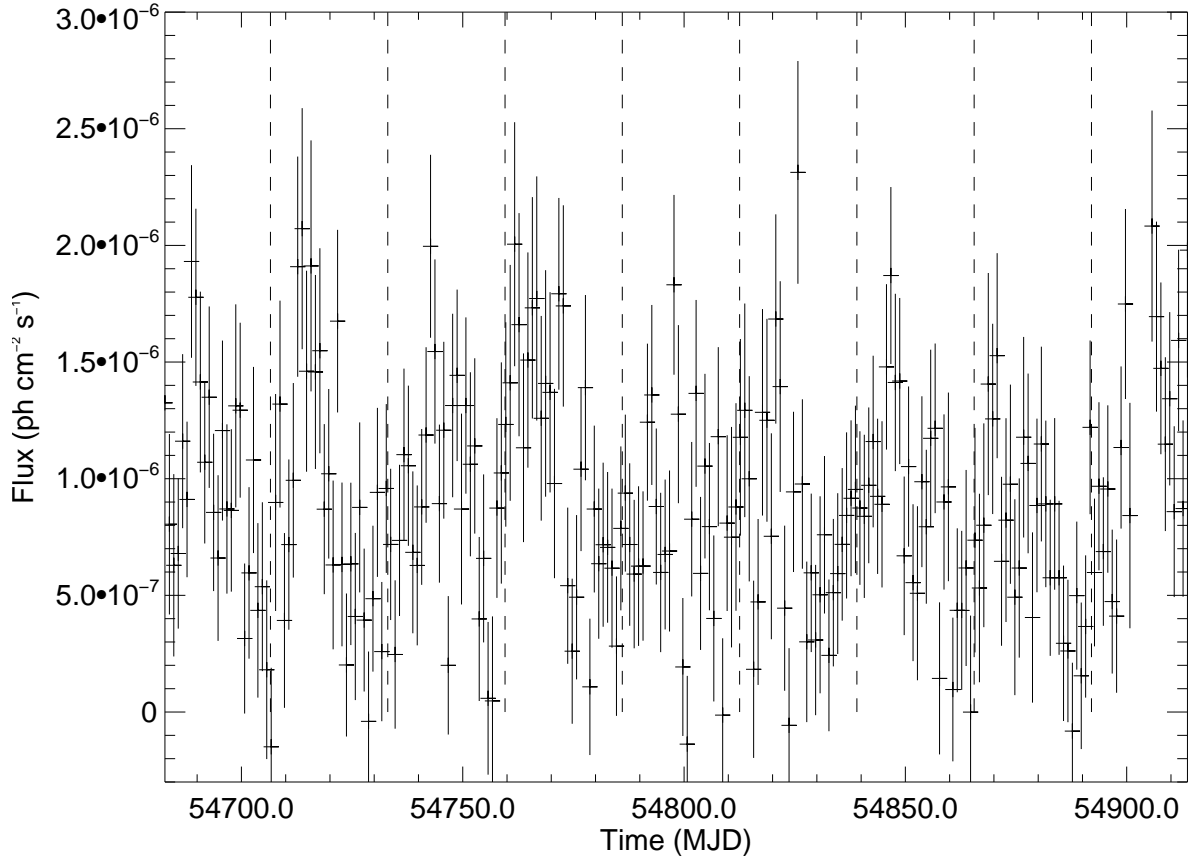}
\caption{The 100 MeV to 20 GeV 1-day binned \emph{Fermi} light curve
  of \lsicap\ covering the period 2008 August 4 through 2009 March 24.
  The vertical lines indicate the zero phase from
  \citet{2002ApJ...575..427G}.}
\label{LC}
\end{figure*}

The {\tt gtlike} likelihood fitting tool was used to perform the
spectral analysis, with "Pass 6 v3" (P6\_V3) instrument response
functions (IRFs); the P6\_V3 IRFs are a post-launch update to address
gamma-ray detection inefficiencies that are correlated with trigger
rate. The 10 degree region around the source was modeled for Galactic
and extragalactic diffuse emission, and included one nearby point
source at (R.A., Dec) of (02$^{\rm h}$23$^{\rm m}$12$^{\rm s}$,
62\degr0\arcmin0\arcsec), too faint to be found in the 3-month Bright
Source List \citep{2009arXiv0902.1559A}. It is important to include
this nearby source in the fitting model because at low energies the
PSF is sufficently wide that despite being $\sim$2.2\degr away the PSF
wings extend across the location of \lsi\ contributing approximately
13\% to the flux at this position.  Simultaneous modelling of this
source accounts for it's contribution to the flux in this region and
any uncertainty is folded into the statistical error of the flux of
\lsi\ found by the likelihood fitting tool.  The 10\degr\ region was
chosen to capture the broad PSF obtained at 100 MeV. An alternate
fitting method using energy-dependent regions of interest was used,
yielding compatible results that were folded into the systematic
errors.
  
The Galactic diffuse emission was modeled using GALPROP, described in
\citet{2004ApJ...613..962S} and \citet{2007Ap&SS.309...35S}, updated
to include recent H I and CO surveys, more accurate decomposition into
Galactocentric rings, and many other improvements, including some from
comparison with LAT data \citep{2009b.Adbo}. The GALPROP run
designation for our model is 54\_59varh7S. The diffuse sources
contribute $\sim$95\% of the observed photons shown in
Fig.~\ref{CountsMap}.

Initially a simple power law, $E^{-\Gamma}$, was fit to the orbital
phase-averaged data yielding a photon index of $\Gamma \sim 2.42$.
However, as indicated in Fig.~\ref{PhaseAveFit}, the energy spectrum
appears to turn over at energies $\sim$6 GeV.  The possibility of an
exponential cutoff was investigated, in the form $E^{-\Gamma}
\exp{[-(E/E_{\rm cutoff})]}$. The chance probability to incorrectly
reject the power law hypothesis was found to be 1.1$\times$10$^{-9}$.
The best fit exponential cutoff returns a test statistic
\citep{1996ApJ...461..396M} significance value of about 4770, or
roughly 70$\sigma$.  The photon index is $\Gamma=$2.21 $\pm$
0.04 (stat) $\pm$ 0.06 (syst); the flux above 100 MeV is (0.82 $\pm$
0.03 (stat) $\pm$ 0.07 (syst)) $\times 10^{-6} $ ph cm$^{-2}$ s$^{-1}$
and the cutoff energy is 6.3 $\pm$ 1.1(stat) $\pm$ 0.4(syst) GeV (see
below for a discussion of systematics). A total of 135,659 photons
were found in the 10\degr\ region. Evaluating the fit parameters, 6467
$\pm$ 80 photons were observed from \lsi\ above 100 MeV.
Fig.~\ref{PhaseAveFit} shows the best fit cutoff power law model as
well as the fluxes fit per energy bin and archival data from MAGIC
\citep{2009ApJ...693..303A} and VERITAS \citep{2008ApJ...679.1427A}.

A number of effects are expected to contribute to the systematic
errors. Primarily, these are uncertainties in the effective area and
energy response of the LAT as well as background contamination. These
are currently estimated by using outlier IRFs that bracket our nominal
ones in effective area. These are defined by envelopes above and below
the P6\_V3 IRFs by linearly connecting differences of (10\%, 5\%,
20\%) at log(E$/$MeV) of (2, 2.75, 4) respectively.  Other potential
sources of systematic effects investigated are: fitting technique;
cuts applied (zenith angle, minimum and maximum energies); and details
of the diffuse modeling.  The systematic errors estimated using the
bracketing IRFs were found to be greater than these additional
effects, hence the bracketing IRF results were quoted for the upper
limits on the systematics.

\subsection{Timing Analysis}

LAT light curves were extracted using aperture photometry.  The LAT
point spread function is strongly energy dependent and, particularly
since \lsi\ is located in the Galactic plane, there is also
significant contribution to the flux within an aperture from diffuse
emission and point sources that depends on the aperture size and the
energy range used.  The aperture and energy band employed were
independently chosen to maximize the signal-to-noise level. The
optimum aperture radius was found to be approximately 2.4\degr\ in the
energy range 100~MeV-20~GeV.  The time resolution of the light curve
was 11,478 s, equal to twice the {\em Fermi} orbital period.Exposures
were calculated using {\tt gtexposure} and used to determine the count
rate in each time bin. In the exposure calculation, the spectral shape
is assumed to be a power-law with a photon index of 2.4.  The 1-day
binned light curve is shown in Fig.~\ref{LC}.  Contributions from the
nearby source and Galactic and extragalactic diffuse backgrounds were
estimated based on the spectral fit and subtracted from the light
curve

\begin{figure}[t]
\includegraphics[width=0.8\linewidth,angle=-90,clip]{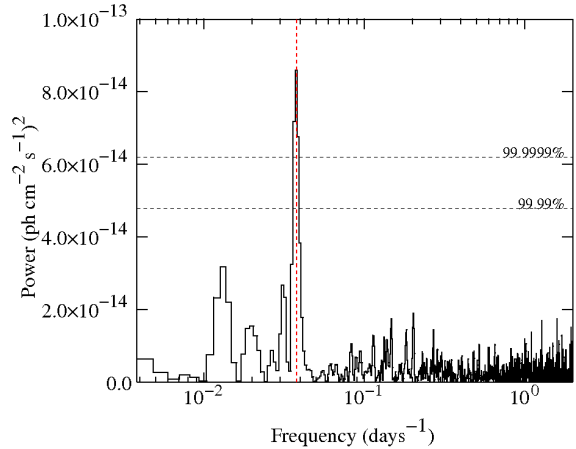}
\caption{Power spectrum of the light curve. The vertical line
  indicates the known orbital period from \citet{2002ApJ...575..427G},
  coinciding with a strong peak in the spectrum, while the horizontal
  lines indicate the marked significance levels.}
\label{Powspec}
\end{figure}

A search was made for periodic modulation by calculating the
periodogram of the light curve
\citep{1976Ap&SS..39..447L,1982ApJ...263..835S}. Since the exposure of
the time bins was variable, the contribution of each time bin to the
power spectrum was weighted based on its relative exposure.  The
periodogram of the unbinned, unsmoothed light curve is shown in
Fig.~\ref{Powspec}.  The vertical line marks the
\citet{2002ApJ...575..427G} orbital period and a highly significant
peak is detected at this period.  The significance levels marked are
for a ``blind'' search with 500 independent frequency steps, however,
the effects of the tuning of the aperture radius and energy range are
not taken into account.  The period and its error from the LAT
observations were estimated using a Monte-Carlo approach: light curves
were simulated using the observed \lsi\ light curve and randomly
shuffling the data points within their statistical errors, assuming
Gaussian statistics.  The corresponding periodogram was then
calculated and the location of the peak at $\sim$26.5 days recorded.
From $\sim$250,000 simulations the distribution of values gives an
estimation of the measured orbital period and its associated error of
26.6 $\pm$ 0.5 days (1$\sigma$).

\begin{figure}[t]
\includegraphics[width=0.8\linewidth,angle=-90,clip]{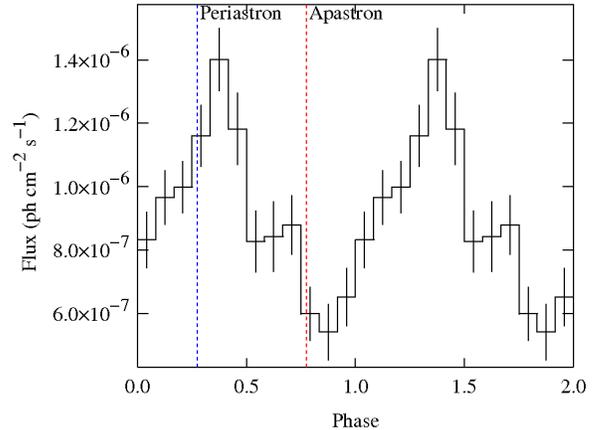}
\caption{Folded light curve of \lsicap\ binned in orbital phase, see
  text for details.  The dashed-lines indicate periastron and apastron
  as given by \citet{2009arXiv0902.4015A}}
\label{FoldedLC}
\end{figure}

The binned LAT light curve folded on the \citet{2002ApJ...575..427G}
period with zero phase at ${\rm MJD}~43,366.2749$
\citep{1979AJ.....84.1030G} is shown in Fig.~\ref{FoldedLC}.  The
folded light curve shows a large modulation amplitude with maximum
flux occurring slightly after periastron passage. The overall light
curve can be fit reasonable well by a simple sine wave, yielding a
reduced $\chi^2_{\nu}$ of 1.4 for 1682 d.o.f.  However, if we use the
known orbital period and ephemeris of the system
\citep{2002ApJ...575..427G} to fit a sine wave to each of the
individual 9 orbits observed then we find that the best fit amplitude
varies between 6.8$\pm$0.9 and 2.2$\pm$0.9 $\times$10$^{-7}$ ph
cm$^{-2}$ s$^{-1}$, which suggests some orbit-to-orbit variability.
 
\subsection{Phase resolved spectral analysis}
The possibility of the spectral shape changing across the orbit was
explored by running {\tt gtlike} fits for phase-folded bins of 0.1
width. The reduced statistics in each phase bin result in a cutoff not
being statistically required to fit the data and so a simple power law
model is used.  There is no significant dependence of photon index on
phase; a fit to a constant value returns a reduced $\chi^2_{\nu}$ of
1.4 for 9 d.o.f, consistent with no variation.
 
\section{Discussion and concluding remarks}

The {\em Fermi} data enable for the first time the detection of a
modulation in GeV gamma rays at the orbital period of a binary
system. The derived period is in excellent agreement with the radio
and optical-based ephemeris \citet{2002ApJ...575..427G}. The COS-B
source 2CG 135+01 is now firmly identified as the gamma-ray
counterpart to \lsi, resolving a 30-year long suspicion that the two
were associated. With the identification originally based on
localization only, the detection of orbital-modulated very high-energy
emission ($>$100 GeV, VHE) from \lsi\ by MAGIC and VERITAS
\citep{2006Sci...312.1771A, 2009ApJ...693..303A, 2008ApJ...679.1427A}
had already provided very strong support in favour of this association.

\lsi\ is detected at a mean flux level above 100 MeV consistent with
that seen by EGRET and {\em AGILE}. Averaged over the orbital
modulation, the source persists as one of the brightest high-energy
gamma-ray sources in the sky over a timescale of decades
\citep[see][Bright Source List, in which this source is the 15$^{\rm
    th}$ brightest]{2009arXiv0902.1559A}.  The folded {\em Fermi}
light curve peaks around phase 0.3, which is compatible with periastron
passage (when the compact object is closest to the Be star) according
to the latest radial velocity studies
\citep{2009arXiv0902.4015A}. This contrasts with the behavior at very
high energies where peak flux occurs at phases 0.6-0.7 and detections
are achieved only at phases ranging from 0.5 to 0.8, before or at
apastron. In X-rays, \lsi\ also appears to peak at phases 0.6-0.7
\citep{1997A&A...320L..25P, 2007A&A...474..575E} whereas the radio
peak occurs over a wide range of phases depending upon a 4-year
super-orbital cycle \citep{2002ApJ...575..427G}.

The average {\em Fermi} and EGRET spectra have compatible power law
indices and fluxes taking into account systematics, but the {\em
  Fermi} spectrum also shows a cutoff at approximately 6 GeV. There is
no evidence for a phase-dependence of the spectral shape and hence,
the index or cutoff energy.  VERITAS reports upper limits during the
only VHE observations that are contemporary with {\em Fermi}, covering
only part of one orbit from phase 0 to 0.75 \citep[up to 2008 November
  9,][]{JaenRef}. The later phases have short exposure
times. Moreover, the past VHE history of the source shows several
non-detections at phases 0.6--0.7 \citep{2008ApJ...679.1427A,
  2009ApJ...693..303A}, perhaps due to variability from one orbital
cycle to the other. The {\em Fermi} light curve displays signs of
orbit-to-orbit variability superposed on the mean behavior, with the
primary peak always around phase 0.3.  Such variability could be
attributed to changing conditions in the Be star wind, affecting the
interaction with the pulsar wind or relativistic jet. Indeed, optical
spectra show evidence for changes in wind emission with the orbit
\citep{1999A&A...351..543Z}.

The obvious radiative process to invoke in the HE and VHE range is
inverse Compton scattering of the abundant stellar photons into
gamma rays by a population of electrons accelerated in the vicinity of
the compact object (e.g. in a relativistic jet or in a pulsar
wind). Then, all else being equal, the peak flux phase is determined
by where the seed photon density is highest and by geometry; favorable
when the high energy electrons are seen behind the star by the
observer, e.g. \citet{2008A&A...477..691D, 2008MNRAS.383..467K,
  2008APh....30..239S}. Superior conjunction is close in phase to
periastron passage in \lsi\ ($\phi_{\rm per}-\phi_{\rm sup}=0.07$ to
0.17 depending on the orbital solution). Hence, having the {\em Fermi}
flux peak close to periastron is consistent with inverse Compton
emission from electrons located close to the compact object.  The
cutoff in the average spectrum could arise due to radiative losses
(because of different accelerating conditions for electrons, because of the
magnetic field amplitude in the relativistic jet or the pulsar
wind along the orbit and/or because of the greater photon
density at periastron), or due to a varying maximum energy for accelerated
electrons or to pair production on stellar photons for gamma rays
above $\approx$50 GeV \citep{2006A&A...456..801D, 2006A&A...459..901S,
  2008A&A...488...37C, 2009ApJ...693.1462S}. In the latter case,
cascade emission might also be seen in the {\em Fermi} range. All
these effects introduce phase-dependent spectral changes.  Hadronic
interactions related to crossings of the Be star's equatorial wind
(disk) could also contribute \citep{2006MNRAS.372.1585C}. This would
provide an independently varying spectral component to explain why the
HE and VHE emission peak at different phases and vary with orbital
cycle. The expectation is that hadronic interactions would result in
two asymmetric peaks in the light curve whose amplitude depends upon
the intercepted matter density during the crossings and occuring at
phases {\em a priori} unrelated to periastron passage but on the
orientation of the orbit of the compact object relative to the Be star
disk.

Continued monitoring by {\em Fermi} combined with dedicated campaigns
by pointed instruments is needed to better constrain spectral
variability and establish the multiwavelength connections: how do
orbit-to-orbit variations compare in different energy ranges? Are
there separate HE and VHE spectral components?

\acknowledgements

The {\em Fermi}-LAT Collaboration acknowledges generous ongoing support from
a number of agencies and institutes that have supported both the
development and the operation of the LAT as well as scientific data
analysis.  These include the National Aeronautics and Space
Administration and the Department of Energy in the United States, the
Commissariat \`a l'Energie Atomique and the Centre National de la
Recherche Scientifique / Institut National de Physique Nucl\'eaire et
de Physique des Particules in France, the Agenzia Spaziale Italiana
and the Istituto Nazionale di Fisica Nucleare in Italy, the Ministry
of Education, Culture, Sports, Science and Technology (MEXT), High
Energy Accelerator Research Organization (KEK) and Japan Aerospace
Exploration Agency (JAXA) in Japan, and the K.~A. Wallenberg
Foundation, the Swedish Research Council and the Swedish National
Space Board in Sweden.

Additional support for science analysis during the operations phase
from the following agencies is also gratefully acknowledged: the
Spanish CSIC and MICINN and the Istituto Nazionale di Astrofisica in
Italy.

We thank the anonymous referee for useful and constructive comments.

{\it Facility:} \facility{Fermi}

\end{document}